\documentclass[10pt]{article}
\usepackage{fullpage}
\usepackage{amsmath}
\usepackage{amssymb}
\usepackage[dvips]{epsfig}
\usepackage{color}

\def\##1{{\bf #1}}
\def\=#1{\underline{\underline #1}}

\def\eps{\epsilon}
\def\epso{\epsilon_0}
\def\muo{\mu_0}
\def\ko{k_0}
\def\lambdao{\lambda_0}
\def\Omegao{\Omega_0}
\def\etao{\eta_0}
\def\.{\mbox{ \tiny{$^\bullet$} }}

\def\ux{\#{u}_x}
\def\uy{\#{u}_y}
\def\uz{\#{u}_z}

\def\un{\#{u}_n}
\def\ut{\#{u}_\tau}
\def\ub{\#{u}_b}

\def\le{\left(}
\def\ri{\right)}
\def\les{\left[}
\def\ris{\right]}
\def\lec{\left\{}
\def\ric{\right\}}

\def\c#1{\cite{#1}}
\def\l#1{\label{#1}}
\def\r#1{(\ref{#1})}

\begin{document}

\begin{center}

{\bf {\LARGE Modeling  chiral sculptured thin films as platforms for
surface--plasmonic--polaritonic optical sensing}}

\vspace{10mm} \large

 Tom G. Mackay\footnote{E--mail: T.Mackay@ed.ac.uk.}\\
{\em School of Mathematics and
   Maxwell Institute for Mathematical Sciences\\
University of Edinburgh, Edinburgh EH9 3JZ, UK}\\
and\\
 {\em NanoMM~---~Nanoengineered Metamaterials Group\\ Department of Engineering Science and Mechanics\\
Pennsylvania State University, University Park, PA 16802--6812,
USA}\\
 \vspace{3mm}
 Akhlesh  Lakhtakia\footnote{E--mail: akhlesh@psu.edu}\\
 {\em NanoMM~---~Nanoengineered Metamaterials Group\\ Department of Engineering Science and Mechanics\\
Pennsylvania State University, University Park, PA 16802--6812, USA}

\normalsize

\vspace{15mm} {\bf Abstract}

\end{center}

\vspace{4mm}

Biomimetic nanoengineered metamaterials  called chiral sculptured thin films (CSTFs) are attractive
platforms for optical sensing because their porosity, morphology and
optical properties can be tailored to order. Furthermore,
their ability to support more than one surface-plasmon-polariton
(SPP) wave at a planar interface with a metal offers functionality
beyond that associated with conventional SPP--based sensors. An
empirical model was constructed to describe SPP-wave propagation guided by
the planar interface of a CSTF~---~infiltrated with a fluid which
supposedly contains analytes to be detected~---~and a metal. The
inverse Bruggeman homogenization formalism was first used to determine the
nanoscale model parameters of the CSTF. These parameters then served as
inputs to the forward Bruggeman homogenization formalism to determine
the reference relative permittivity dyadic of the infiltrated CSTF. By solving
the coresponding boundary-value problem for a modified Kretschmann
configuration, the characteristics of the multiple SPP modes at the planar interface were
investigated as functions of the refractive index of the fluid
infiltrating the CSTF and the rise angle of the CSTF. The SPP
sensitivities thereby revealed bode well
 for the implementation of fluid--infiltrated CSTFs
 as SPP--based optical sensors.

 \vspace{5mm}

\noindent {\bf Keywords:} Bruggeman homogenization formalism,
surface plasmon polariton,  chiral sculptured thin film

\section{Introduction}

Engineered biomimicry till date has two major components:
whereas {\it bioinspiration} leads to the same outcome as
a biological activity, biomimetics is the reproduction of a
natural functionality by copying relevant attributes of a biological organism.
Bioinspiration is exemplified by  aeroplanes that emulate the flight of birds
and insects, biomimetics by the hook-and-loop fasteners whose structure
emulates that of the burrs of some plants.

Another example of biomimetics  also has roots in botany \cite{Coll,Sonin,SDS}.
Cholesterol is found
in biological cells containing a nucleus. Benzoic acid is the principal component of
a resin obtained from the bark of certain species of trees. The discovery of
distinct temperature-dependent color effects in cholesteryl benzoate (C$_{34}$H$_{50}$O$_2$), an ester of cholesterol and benzoic acid,
was reported in 1888 by Friedrich Reinitzer, an Austrian botanist \cite{Reinitzer,Kelker}
Within 15 years, the
terms {\it liquid crystal} and {\it flowing crystal} for cholesteryl benzoate and similar materials   had become well established \cite{Sonin,SDS}
due to the relentless research efforts  of Otto Lehmann, with terminological clarity subsequently established in 1922 by Georges
Friedel \cite{Friedel}.

Cholesteric liquid crystals (CLCs) comprise aciculate molecules dispersed on a stack of closely spaced parallel planes. The molecules
lying on a specific plane have an average orientation. This orientation progressively changes from plane to plane, the ensemble
of orientations describing a
helix in the thickness direction. The helix can be either left handed or
right handed, just like curling tendrils that facilitate nutation by creepers and vines \cite{Darwin}.
Given the presence of cholesterol in plant cells, the helical structure with striking optical consequences has been
found in leaves of certain species \cite{GL1996}.
When circularly polarized light of free-space wavelength in a certain regime falls
normally on a CLC, the reflectance is high if the handednesses of the CLC and the incident light are
the same, but is low otherwise---a phenomenon that has found much use over the last three decades \cite{Fergason,Jac92}. Parenthetically
but remarkably, a coarse version of the cholesteric structure had been deduced two decades
prior to Reinitzer's discovery by Ernst Reusch \cite{Reusch}, with surprising optical effects revealed a
century later \cite{Joly1986,Ambi}. However, Reusch's work was inspired not by a botanical specimen but a mineral (mica).

Much before CLCs had not yet reached their current technological prominence, in 1959 Young and Kowal \cite{YK1959} reported the fabrication
of what they called a ``helically evaporated film" or a ``helically deposited film". Furthermore, they stated that this
film ``would resemble a Solc filter of the fan type, were the retardation plates of the Solc [{\it sic}] filter to approach
zero thickness (and infinite number) while maintaining the total thickness of the filter constant." A year later,
Dawson and Young \cite{DY1960} related the structure of the Sol\u{c} filter of the fan type to
that of ``Reusch rotators", with their own films
to be obtained therefrom ``by allowing the number of elements to approach infinity". They had thus
engineered a solid-state analog of CLCs.

Although hugely significant, the 1959 paper of Young and Kowal \cite{YK1959} became obscure,
gathering just 5 citations until and including 1996, according to the Web of Science\texttrademark, out of
 a total of 85 at the time of this writing.\footnote{The 1960 paper of Dawson and Young \cite{DY1960} fared worse in achieving
 a celebrity status: it will probably receive its 7th
 citation in the Web of Science when this paper is published.} The underlying concept was resuscitated in the early 1990s, with the emergence of
 sculptured thin films (STFs) \cite{LM1,LMBR}, which led to renewed interest in the 1959 paper.

STFs exemplify {\it nanoengineered biomimetic metamaterials}. An STF is an assembly of parallel
nanowires  whose bent and twisted shapes are
engineered via dynamic manipulation during a physical vapor deposition process involving the production
and aggregation of 1--to--3-nm clusters of atoms \cite{STF_Book}. As STFs can be
multifunctional \cite{LMBR,STF_Book,LDHX}, it is appropriate to call them metamaterials \cite{Walser,LMopn}.

The nanoscale control over the
morphology of STFs has meant that their optical properties can be controlled on subwavelength scales \cite{STF_Book}. Among
a bevy of optical applications reported, the most prominent invokes the
circular Bragg phenomenon exhibited by a chiral STF (CSTF). Comprising helical nanowires, CSTFs have been designed and verified
to function as
circular polarization filters \c{Wu} of the  wideband \cite{FC2004,PSH2008}, narrowband \cite{Hod00}, and multiband \cite{PSH2008} varieties.
Spectral shifts of narrowband filters due to fluid infiltration have also been modeled
\cite{ML_IEEEPJ} and demonstrated \cite{LMSWH01,Horn}  for optical sensing. Very significantly, the planar interface
of a CSTF and a metal has been theoretically shown \cite{Polo_PRSA,Polo_JOSAA}
to support the propagation of multiple surface-plasmon-polariton (SPP) waves---all
of the same frequency but different phase speed, attenuation rate, and field configuration---and preliminary experimental evidence \cite{DPL}
is promising.

From a quantum mechanical viewpoint, an SPP is a
quasi-particle which travels along the interface
of a metal and a dielectric material, arising from the
interaction of photons in the dielectric material and electrons in
the metal \c{Felbacq}. A train of SPPs constitutes a SPP wave in
classical language. Since SPP waves are  acutely sensitive
to the morphological and constitutive properties of the materials on either side of the interface to
which they are localized, they have been widely exploited in
chemical and biological sensing applications \c{Homola2003}. Indeed,
SPP--based biosensors are at the forefront of optical label-free and
real-time detection of analytes related to medical diagnostics,
environmental monitoring and food safety \c{Fan,Homola2008,AZL2,Scarano}.

Essentially, an SPP--based sensor detects changes in refractive
index arising in the partnering dielectric material, as follows.
Suppose that a light wave excites an SPP wave which propagates along
the interface of a metal and a dielectric material. The evanescent
field of the SPP wave probes the dielectric material. Small changes
in the refractive index of the dielectric material~---~for example,
brought about by analyte molecules binding with biorecognition
elements immobilized in the vicinity of the interface~---~results in
detectable changes to the light wave coupled to the SPP wave, such
as changes in angle and wavelength of excitation, intensity and
phase \c{Homola2003,AZL2}.

A key design feature of SPP--based sensors is the porosity of the partnering
dielectric material, which allows analyte molecules to access the
vicinity of the metal/dielectric interface. Therefore, STFs,
 whose porosity can be engineered to a high
degree by controlling the physical vapor deposition process used for
their manufacture \c{STF_Book}, are promising candidates as
platforms for SPP--based sensors. Recently, we reported on the
sensitivity of SPP waves excited at the interface of a metal and  a
 certain type of STF known as a columnar thin film (CTF),
to changes in the refractive index of a fluid which infiltrates the
void regions of the CTF \c{ML_PNFA}. Our numerical studies revealed
that SPP waves propagate at a lower
  phase speed and with a shorter propagation length, if the infiltrating
  fluid has a larger refractive index.
Furthermore, the angle of incidence required to excite an SPP wave
  in a Kretschmann configuration \c{Homola2003,AZL2}  increases as the  refractive index of the fluid increases.

We consider in the following sections the potential for a metal/CSTF
interface to act  as a platform for a SPP--based
sensor. A CSTF may be regarded as  a periodically nonhomogeneous
continuum, which  displays orthorhombic symmetry locally but is
structurally chiral from a global perspective \cite[Chap.
9]{STF_Book}. The periodic nonhomogeneity of CSTFs engenders an especially
interesting property not shared by homogeneous materials \cite{PLreview}: a CSTF can
support more than one mode of SPP-wave propagation at its
interface with a metal \c{Polo_PRSA,Polo_JOSAA,DPL}, thereby
opening up the possibility of simultaneous detection of more than
one type of analyte molecule. We note  that CSTFs also offer  an
alternative sensing route, based on spectral shifts of the circular
Bragg phenomenon brought about by the infiltrating fluid containing analytes  \c{ML_IEEEPJ}, that could conceivably
be harnessed in parallel with SPP--based detection to further extend
the efficacy of the optical biosensing device.

In the notation adopted, vectors and column vectors are in boldface
with the later enclosed in square brackets. Dyadics and matrixes are
double underlined with the later enclosed in square brackets; a
superscript `T' denotes the transpose. The Cartesian unit vectors
are written as $\ux$, $\uy$ and $\uz$. The free-space wavenumber,
the free-space wavelength, and the intrinsic impedance of free space
are given by $\ko=\omega\sqrt{\epso\muo}$, $\lambdao=2\pi/\ko$ and
$\etao=\sqrt{\muo/\epso}$, respectively, with $\muo$ and $\epso$
being  the permeability and permittivity of free space. An
$\exp(-i\omega t)$ time-dependence is implicit, with $\omega$
denoting the angular frequency and $i = \sqrt{-1}$.

\section{Theory}
As a realistic setup for launching SPP waves along the planar
interface of a metal film and a CSTF, we consider the modification
\c{Lakh_oc} to the standard Kretschmann configuration \c{Kret}
illustrated schematically in Fig.~\ref{fig1}. The regions $z
\leq 0$ and $z \geq L_\Sigma$ are assumed to be occupied by homogeneous,
isotropic, nondissipative, dielectric materials with relative
permittivity scalars $\eps_{d}$ and $\eps_{\ell}=n_\ell^2$,
respectively. A CSTF~---~infiltrated by a fluid of refractive index
$n_\ell$~---~occupies the laminar region $L_m \leq z \leq L_\Sigma$,
while the laminar region $0 \leq z \leq L_m$ is occupied by a metal
with relative permittivity $\eps_m$.

\subsection{Constitutive and morphological parameters of CSTF}

A CSTF comprises an array of parallel helical nanowires \cite{STF_Book}. It
may  been grown on a planar substrate,  lying parallel to the plane
$z=0$ say, by the deposition of an evaporated bulk material. On
rotating the
 substrate  about
the $z$ axis at a uniform angular speed throughout the deposition
process, helical nanowires grow along the $z$ direction, with the
rise angle of each  nanowire, relative to the $xy$ plane, being
denoted by $\chi$.
 The deposited material is
assumed to be an isotropic dielectric material of refractive index
$n_s$. Significantly, $n_s$ can be different from the refractive
index of the bulk material that was evaporated
\c{MTR1976,BMYVM,WRL03}.

The helical shape of each nanowire of a CSTF  can be viewed as a
string of highly elongated ellipsoidal inclusions, wound end-to-end
around the $z$ axis  \cite{Sherwin,Lakh_Opt}.  The  shape dyadic
\begin{equation}
 \un \, \un + \gamma_\tau \, \ut \, \ut + \gamma_b \, \ub \,
\ub ,
\end{equation}
wherein the normal, tangential, and binormal basis vectors are given
as
\begin{equation}
 \un = - \ux \, \sin \chi + \uz \, \cos \chi\,,\quad
 \ut =  \ux \, \cos \chi + \uz \, \sin \chi\,,\quad
\ub = - \uy
\,,
\end{equation}
prescribes the surface of each ellipsoidal inclusion. An aciculate
shape is imposed on the inclusions by selecting the shape parameters
$\gamma_{b} \gtrsim 1$ and $\gamma_\tau \gg 1$. Increasing
$\gamma_\tau$ beyond 10 does not give rise to significant effects
for slender inclusions \cite{Lakh_Opt}. Accordingly, for the
numerical results which follow in \S\ref{Numerica},
  $\gamma_\tau = 15$ was chosen.
The proportion of a CSTF's total volume occupied by helical
nanowires is $f \in \le 0, 1 \ri $; equivalently,
 the  volume fraction  of the CSTF not occupied by nanowires is $1 -
f$.

The relative permittivity dyadic
\begin{equation}
\=\eps_{\,stf} = {\=S}_{\,z} \le h  \frac{\pi (z-L_m)}{\Omega} \ri \.
{\=S}_{\,y} \le \chi \ri \. \=\eps^{ref}_{\,stf} \. {\=S}^T_{\,y}
\le \chi \ri \. {\=S}^T_{\,z} \le h \frac{\pi (z-L_m)}{\Omega} \ri\,,\quad
L_m \leq z \leq L_\Sigma\,,
\l{eps1_dyadic}
\end{equation}
characterizes the CSTF at length scales much greater than the
nanoscale. Herein  the handedness parameter $h = + 1 (-1)$ for a
structurally right (left)-handed CSTF;
 the
rotation dyadics
\begin{equation}
\left.
\begin{array}{l}
{\=S}_{\,y} \le \chi \ri = \#u_y\, \#u_y + \le \#u_x\, \#u_x +
\#u_z\, \#u_z \ri \cos \chi + \le \#u_z\, \#u_x - \#u_x\, \#u_z \ri
\sin \chi \vspace{4pt} \\
{\=S}_{\,z} \le \sigma \ri =
 \#u_z\, \#u_z +
\le \#u_x\, \#u_x + \#u_y\, \#u_y \ri \cos  \sigma  + \le \#u_y\,
\#u_x - \#u_x\, \#u_y \ri \sin  \sigma
\end{array}
\right\}\,;
\end{equation}
and  $2 \Omega$ is the structural period. We take $l_{stf} =
(L_\Sigma - L_m)/2 \Omega$ to be a positive--valued integer; i.e.,
the CSTF contains a whole number of structural periods.
 The reference
relative permittivity dyadic $\=\eps^{ref}_{\,stf}$ has the
orthorhombic form
\begin{equation}
\=\eps^{ref}_{\,stf} = \eps_{a \nu}   \,\un\,\un +\eps_{b
\nu}\,\ut\,\ut \, +\,\eps_{c \nu}\,\ub\,\ub\, , \l{eps1_ref_dyadic}
\end{equation}
where $\nu =1$ for a CSTF in which the void regions are vacuous
(i.e., an uninfiltrated CSTF) and $\nu =2$ for a CSTF in which the
void regions are filled with a fluid of refractive index $n_\ell$.

In principle,  the relative permittivity parameters $\lec \eps_{a1},
\eps_{b1}, \eps_{c1} \ric$ of an uninfiltrated CSTF are measurable.
However, in the absence of  suitable measured data  for CSTFs,
recent numerical studies have used values of $\lec \eps_{a1},
\eps_{b1}, \eps_{c1} \ric$ measured for related CTFs. The nanoscale
model parameters $\lec n_s, f, \gamma_b \ric$~---~which are not
readily determined  by experimental means~---~can be determined from
a knowledge of $\lec \eps_{a1}, \eps_{b1}, \eps_{c1} \ric$ by
applying the inverse  Bruggeman
 homogenization formalism \c{ML_inverse_homog}. Once $\lec n_s, f, \gamma_b
\ric$ have been found, they can be combined with $\lec
n_\ell,\gamma_\tau\ric$ in order to determine the relative
permittivity parameters $\lec \eps_{a2}, \eps_{b2}, \eps_{c2} \ric$
for the infiltrated CSTF, by applying the Bruggeman homogenization
formalism  in its usual forward sense \c{ML_IEEEPJ,Lakh_Opt}.

\subsection{Boundary-value problem}

The  essence of the sensing mechanism in the modified Kretschmann
configuration is described by the following boundary-value problem \cite{Lakh_oc}.
An arbitrarily polarized plane wave is launched in the half-space $z
< 0$ towards the metal layer. We suppose that its wavevector lies in
the $xz$ plane, making an angle $\theta_{inc} \in \les 0, \pi/2 \ri$
relative to the $+z$ axis. This incident plane-wave gives rise to  a
reflected plane wave in the half-space $z < 0$ and a transmitted
plane wave in the half-space $z
> L_\Sigma$. Thus, the total electric field phasor in the half-space $z < 0$ may
be expressed as
\begin{eqnarray}
\nonumber
 \#E (\#r) &=& \les \, a_s \#u_y + a_p \#p_+ (\theta_{inc})  \,\ris\,\exp \le i \kappa x \ri \, \exp
\le i \ko \sqrt{\eps_d} \,  z \cos \theta_{inc} \ri
\\
&& + \les \, r_s \#u_y + r_p \#p_- (\theta_{inc})  \,\ris\,\exp \le
i \kappa x \ri \, \exp \le - i \ko \sqrt{\eps_d} \,  z \cos
\theta_{inc} \ri  , \qquad
 z < 0,
\end{eqnarray}
while that in the half-space $z > L_\Sigma$ may be expressed as
\begin{equation}
\#E (\#r) = \les \, t_s \#u_y + t_p \#p_+ (\theta_{tr}) \,\ris\,\exp
\le i \kappa x \ri \, \exp \les i \ko n_\ell \le z - L_\Sigma \ri
\cos \theta_{tr} \ris,
 \qquad z > L_\Sigma,
\end{equation}
wherein $\#p_\pm (\theta)  = \mp \#u_x \cos \theta + \#u_z \sin
\theta$, $\kappa=\ko \sqrt{\eps_d} \, \sin \theta_{inc}$ and the
angle of transmission $\theta_{tr}$ satisfies the law of Ibn Sahl \cite{STF_Book} as follows:
\begin{equation}
\sqrt{\eps_{d}} \, \sin \theta_{inc} = n_{\ell} \, \sin \theta_{tr}.
\end{equation}

By solving the related boundary-value problem, the  complex--valued
reflection and transmission amplitudes, namely $r_{s}$, $r_{p}$,
$t_{s}$ and
 $t_{p}$,
are related to the  corresponding amplitudes $a_s$ and $a_p$ of the
$s-$ and $p-$polarized components of the incident plane wave. This
standard procedure  yields the algebraic relation \c{Polo_PRSA}
\begin{equation}
\les
\begin{array}{c}
t_s \\ t_p \\ 0 \\ 0 \end{array} \ris = \les \,\=K (n_\ell^2,
\theta_{tr}) \,\ris^{-1} \cdot \les \,\=Q\,\ris^{l_{stf}} \cdot
\exp \le i \les\, \=P_{\,m}\, \ris L_m \ri \cdot \les \,\=K (\eps_d,
\theta_{inc}) \,\ris \cdot \les
\begin{array}{c}
a_s \\ a_p \\ r_s \\ r_p \end{array} \ris, \l{bvp}
\end{equation}
wherein the 4$\times$4 matrixes
\begin{equation}
\les \,\#K (\eps, \theta) \,\ris = \les
\begin{array}{cccc}
0 & - \cos \theta & 0 & \cos \theta \\ 1 &0 &1 & 0\\
- \le \sqrt{\eps} \, \cos \theta \ri/\etao &0&  \le \sqrt{\eps} \,
\cos \theta \ri/\etao & 0
\\ 0 & - \sqrt{\eps }/\etao &0 & -\sqrt{\eps }/\etao
\end{array} \ris,
\end{equation}
\begin{equation}
\les \, \#P_{\,m} \, \ris = \les
\begin{array}{cccc}
0 &0 & 0 & \omega \muo - \le \kappa^2/\omega \epso \eps_m
\ri \\
0 &0 & - \omega \muo &0 \\
0 &- \omega \epso \eps_m + \le \kappa^2/\omega \muo \ri& 0
&0 \\
\omega \epso \eps_m  &0&0&0
\end{array} \ris\,.
\end{equation}
The  transfer matrix $\les \,\=Q \,\ris$ of one structural period of the CSTF and its evaluation are
comprehensively described elsewhere \c{STF_Book,Polo_PRSA}. From
\r{bvp}, we may write
\begin{equation}
\les \begin{array}{c} r_s \vspace{4pt} \\
r_p \end{array} \ris = \les \begin{array}{cc} r_{ss}  & r_{sp}
\vspace{4pt} \\ r_{ps} & r_{pp}
\end{array} \ris
 \les \begin{array}{c} a_s \vspace{4pt} \\
a_p \end{array} \ris,  \qquad \quad
\les \begin{array}{c} t_s \vspace{4pt} \\
t_p \end{array} \ris = \les \begin{array}{cc} t_{ss}  & t_{sp}
\vspace{4pt} \\ t_{ps} & t_{pp} \end{array} \ris
\les \begin{array}{c} a_s \vspace{4pt} \\
a_p \end{array} \ris,
\end{equation}
thereby introducing the reflection coefficients $r_{ss,sp,ps,pp}$
and transmission coefficients $t_{ss,sp,ps,pp}$. The square
magnitude of a reflection   coefficient delivers the
corresponding reflectance; i.e.,
\begin{equation}
R_{\alpha \beta}
= \left| r_{\alpha \beta} \right|^2\,,\quad \alpha \in \lec s,
p\ric\,,\quad   \beta \in \lec s,
p\ric\,.
\end{equation} The four transmittances are  defined as
follows:
\begin{equation}
T_{\alpha \beta} = \frac{n_{\ell} \; \mbox{Re} \, \les \, \cos \theta_{tr} \,
\ris}{\sqrt{\eps_{d}} \, \cos \theta_{inc}}\,
 \left|
t_{\alpha \beta} \right|^2\,,\quad \alpha \in \lec s,
p\ric\,,\quad   \beta \in \lec s,
p\ric\,.
\end{equation}

The absorbances for $p-$ and $s-$polarization states of the incident plane wave, namely
\begin{equation}
\left.
\begin{array}{l}
A_p = 1 - \le R_{pp} + R_{sp} +  T_{pp} + T_{sp} \ri  \vspace{6pt}\\
A_s = 1 - \le R_{ss} + R_{ps} +  T_{ss} + T_{ps} \ri
\end{array}
\right\},
\end{equation}
play a key role in identifying SPP waves. A sharp high peak in the
graph of absorbance versus $\theta_{inc}$, arising at $\theta_{inc}
= \tilde{\theta}^{(\tau)}_{inc}$ say, is a signature of SPP
excitation at the $z=L_m$ interface provided that
\begin{itemize}
\item[ (i)] $\tilde{\theta}^{(\tau)}_{inc}$ is insensitive to changes
in the CSTF's thickness  and \item[ (ii)] all eigenvalues of the
matrix $\les\, \tilde{\= Q} \,\ris$, defined by \c{Motyka_II,Part4}
\begin{equation}
\les\,  {\= Q} \,\ris =\exp\lec i2\Omega\les\, \tilde{\= Q}
\,\ris\ric\,,
\end{equation}
 evaluated  at
$\tilde{\theta}^{(\tau)}_{inc}$ have non-zero imaginary parts.
\end{itemize}

\section{Numerical results and discussion} \label{Numerica}

The helical morphology of a CSTF is nanoengineered by obliquely directing
a collimated  vapor flux  in high vacuum towards a planar substrate that rotates about a fixed axis
at a constant rate \cite{STF_Book}. The angle $\chi_v$ between the average direction of the vapor flux and the substrate plane
determines the dielectric properties of the  CSTF.
In view of the absence of suitable experimental data for CSTFs, for
our numerical studies  we chose the relative permittivity parameters
\begin{equation}
\left.
\begin{array}{l}
\eps_{a1} = \displaystyle{\les 1.0443 + 2.7394 \le \frac{2
\chi_v}{\pi} \ri - 1.3697
\le \frac{2 \chi_v}{\pi} \ri^2 \ris^2} \vspace{6pt} \\
\eps_{b1} = \displaystyle{ \les 1.6765 + 1.5649 \le \frac{2
\chi_v}{\pi} \ri - 0.7825 \le \frac{2 \chi_v}{\pi} \ri^2 \ris^2}
 \vspace{6pt} \\
\eps_{c1} = \displaystyle{ \les 1.3586 + 2.1109 \le \frac{2
\chi_v}{\pi} \ri - 1.0554 \le \frac{2 \chi_v}{\pi} \ri^2 \ris^2}
\end{array}
\right\} \l{tio1}
\end{equation}
with
\begin{equation}
\tan \chi = 2.8818 \, \tan \chi_v, \l{tio2}
\end{equation}
which were determined by experimental measurements at a free--space
wavelength of 633 nm on a  CTF
 made from patinal${}^{\mbox{\textregistered}}$
titanium oxide
\c{HWH_AO,Chiadini1}. Values for the corresponding nanoscale model
parameters $\lec n_s, f, \gamma_b \ric$,  as computed  using the
inverse Bruggeman homogenization formalism \c{ML_inverse_homog}, are
provided in Table~\ref{tab1} for the vapor flux angles  $\chi_v =
15^\circ$, $30^\circ$, and $60^\circ$. Furthermore, we set $h=+1$ and
normalized the CSTF's structural half-period $\Omega$ relative to $\Omegao =
197$ nm.
 The metal was taken to be aluminum with relative permittivity
$\eps_m = -56 + 21i$ at $\lambdao = 633$ nm, and the thickness of
the metal film $L_m = 15$ nm. The relative permittivity $\eps_d =
6.656$ which is typical of zinc selenide.

The  parameter values chosen are consistent with those chosen for an
earlier study involving an uninfiltrated CSTF ($n_\ell=1$)
\c{Polo_PRSA}, thereby allowing direct comparisons to be made.
Extrapolating from this earlier study, we expect there to be up to
five modes of SPP-wave propagation at the planar metal/CSTF
interface, depending upon the value of the ratio $\Omegao/\Omega$.
The manifestation of these SPP modes in graphs of absorbance versus
angle of incidence was tracked from the earlier study \c{Polo_PRSA}
to the present study by continuously varying the refractive index
$n_\ell$. Further confirmation of the SPP status of the absorbance
peaks was provided by checking that the corresponding eigenvalues of
the transfer matrix $\les \, \tilde{\=Q} \,\ris$ have non-zero
imaginary parts, per \c{Motyka_II,Part4}.

We begin our presentation of numerical results with the graphs of
  absorbance for incident light of  $s-$ and $p-$polarization states versus $\theta_{inc}$ in Fig.~\ref{fig2}. Here $n_\ell
= 1.2$, $l_{stf} = 2$, $\Omegao / \Omega = 1.5$ and $\chi_v =
20^\circ$. The $A_p$ peaks at $\theta_{inc} = 52.7^\circ$ and
$33.3^\circ$ represent the SPP modes 1 and 2, respectively (in the
terminology of Polo and Lakhtakia \c{Polo_PRSA}).

The non-SPP $A_p$ peaks and the $A_s$ peaks which appear in
Fig.~\ref{fig2} indicate waveguide modes \cite{Motyka1} which must depend upon the
thickness of the CSTF. To demonstrate this, in Fig.~\ref{fig3} the
graphs of  Fig.~\ref{fig2} are reproduced  except that here $l_{stf}
= 3$. We see that the $A_p$ peaks corresponding to the SPP modes 1
and 2 remain fixed in position but the other $A_p$ and $A_s$ peaks
have moved relative to their positions in Fig.~\ref{fig2}.

There are no $A_s$ peaks corresponding to the two SPP $A_p$ peaks in
Figs.~\ref{fig2} and \ref{fig3}. However, by considerably increasing
the thickness of the CSTF, evidence of a $A_s$ SPP emerges.
Fig.~\ref{fig4} is as Figs.~\ref{fig2} and \ref{fig3} except that
here $l_{stf} = 20$ and we focus on the angle of incidence
 range $32.5^\circ < \theta_{inc} < 38.5^\circ$.
 We see that a
$A_s$ peak has emerged at $\theta_{inc} = 33.8^\circ$; i.e., at the
same $\theta_{inc}$ position as the  mode 2 SPP $A_p$ peak. A
corresponding $A_s$ peak was reported for the case of an
uninfiltrated CSTF \c{Polo_PRSA}. The large thickness needed is due to the
slow decay rate of the electromagnetic fields in the thickness direction \c{Polo_JOSAA}.

The SPP modes 3--5 only exist only for values of $\Omegao/\Omega$
lower than that considered in Figs.~\ref{fig2}--\ref{fig4}.
 In Fig.~\ref{fig5} the $A_p$
peaks corresponding to SPP modes 3 and 4 can be seen at
$\theta_{inc} = 46.1^\circ$ and $\theta_{inc} = 43.3^\circ$,
respectively, for $\Omegao/\Omega = 0.2$ in the plot of absorbance
versus angle of incidence. Also, the $A_p$ peak corresponding to SPP
mode 5
 can be seen  at $\theta_{inc} = 44.3^\circ$ for $\Omegao/\Omega = 0.13$
 in the plot of absorbance versus $\theta_{inc}$  in Fig.~\ref{fig6}.
As in Figs.~\ref{fig2}--\ref{fig4}, the non--SPP peaks which appear
in Figs.~\ref{fig5} and ~\ref{fig6} represent waveguide modes, as
has been confirmed by additional computations (not presented here)
for different CSTF thicknesses.

The values of $\theta_{inc}$ at which the SPP peaks appear in the
graphs of absorbance are sensitive to both the refractive index
$n_\ell$ of the fluid infiltrating the CSTF and the vapor incidence
angle $\chi_v$.  The value of $\theta_{inc}$ at which the absorbance
peak arises for the SPP mode $\tau$~---~let us denote this value as
$\tilde{\theta}^{(\tau)}_{inc}$ at a specific value of
$n_\ell$~---~is plotted versus $n_\ell \in \le 1, 1.5 \ri$ and
$\chi_v\in \le 15^\circ, 60^\circ \ri$ for $\tau =1 $, and versus
$n_\ell \in \le 1, 1.5 \ri$ and  $\chi_v\in \le 15^\circ, 30^\circ
\ri$ for $\tau =2$, in Fig.~\ref{fig7}. Here $l_{stf} = 2$ and
$\Omegao / \Omega = 0.6$. It may be observed that
$\tilde{\theta}^{(1)}_{inc}$ uniformly increases as $n_\ell$
increases and as $\chi_v$ increases. A similar trend is exhibited by
$\tilde{\theta}^{(2)}_{inc}$. The corresponding plots for
$\tilde{\theta}^{(3)}_{inc}$, $\tilde{\theta}^{(4)}_{inc}$ and
$\tilde{\theta}^{(5)}_{inc}$ (not provided here) also follow the
same general trends as those displayed in Fig.~\ref{fig7}. We note
that for $\chi_v \gtrapprox 65^\circ$ and $n_\ell \gtrapprox 1.3$,
the SPP mode 1 vanishes as $\tilde{\theta}^{(1)}_{inc}$ approaches
$90^\circ$. Furthermore, as compared to the SPP mode 1, the SPP mode
2 exists only for a  smaller  $\chi_v$-range. Indeed, as compared to
the SPP mode 2, the SPP modes 3--5  exist for even  smaller
$\chi_v$-ranges.
 The general trends displayed  in
Fig.~\ref{fig7} are the same as those reported for an analogous
study based on an infiltrated CTF \c{ML_PNFA}.

For optical-sensing applications, a feature of practical
significance is the shape of the SPP peaks in  graphs of absorbance
versus angle of incidence. For the  SPP mode 1 represented in
Fig.~\ref{fig7}, the sharpness of this $A_p$  peak, as gauged by the
second derivative $d^2 \le \left| A_p
 \right|^2 \ri / d
 \theta^2_{inc}$    evaluated at $\theta_{inc} =
 \tilde{\theta}^{(1)}_{inc}$, is plotted in Fig.~\ref{fig8}
  against $n_\ell \in \le 1, 1.5 \ri$ and $\chi_v \in \le 15^\circ,
60^\circ \ri$.
 The sharpness of the selected
peak decreases  as $n_\ell$ increases from 1.0 and  as
$\chi_v$ increases from $15^\circ$. The most dramatic changes in
sharpness occur at when both  $n_\ell$ and  $\chi_v$ have low
values. The
 trends in Fig.~\ref{fig8} for the SPP mode 1 are qualitatively similar to  those reported for an
analogous study based on an infiltrated CTF \c{ML_PNFA}, over the
range $\chi_v \in \le 15^\circ, 60^\circ \ri$. Also provided in
Fig.~\ref{fig8} is the corresponding plot of $\log \les d^2 \le
\left| A_p
 \right|^2 \ri / d
 \theta^2_{inc} \ris$    evaluated at $\theta_{inc} =
 \tilde{\theta}^{(2)}_{inc}$. The $A_p$ peak for the SPP mode 2 is clearly
 much sharper than the peak for the SPP mode 1, across the entire range of $n_\ell$ values and $\chi_v$ values  considered.
  Similarly, the $A_p$ peaks for the SPP
 modes
 3--5  were found to be much sharper than the peak for the SPP mode 1.

Let us now further consider the sensitivity of the SPP peaks in the
graphs of absorbance versus angle of incidence
 to both
$n_\ell$ and $\chi_v$.  As in Figs.~\ref{fig7} and \ref{fig8}, we
fix
 $l_{stf} = 2$ and $\Omegao / \Omega = 0.6$.
The
 figure of merit (in degree/RIU\footnote{RIU = refractive-index unit})
\begin{equation}
\rho^{(\tau)} = \frac{\tilde{\theta}^{(\tau)}_{inc} (n_\ell) -
\tilde{\theta}^{(\tau)}_{inc} (1.0)}{n_\ell - 1.0}\,,\quad
\tau\in\lec1,2,3,4,5\ric\,,
\end{equation}
wherein $\tilde{\theta}^{(\tau)}_{inc}$ is expressed as a function
of $ n_\ell$, is a measure of sensitivity for the SPP mode $\tau$.
 Graphs of $\rho^{(1)}$ versus  $n_\ell \in (1.0, 1.5)$ are provided in
Fig.~\ref{fig9} for $\chi_v \in
 \lec 15^\circ, 30^\circ, 60^\circ \ric$. Also,
$\rho^{(2)}$ is plotted versus  $n_\ell \in (1.0, 1.5)$  for $\chi_v
\in
 \lec 15^\circ, 21^\circ, 30^\circ \ric$.
Generally,
 $\tilde{\theta}^{(1)}_{inc} (n_\ell)$ is most sensitive to
changes in
 $n_\ell$ when $\chi_v$ is small and $n_\ell$ is large, in keeping
 with  an
analogous study based on an infiltrated CTF \c{ML_PNFA}. A similar
trend is followed by  $\tilde{\theta}^{(2)}_{inc} (n_\ell)$---and by
$\tilde{\theta}^{(\tau)}_{inc} (n_\ell)$, $\tau\in\lec3,4,5\ric$,
which are not represented in Fig.~\ref{fig9}. Furthermore, the
magnitudes of $\rho^{(\tau)}$ for all $\tau \in \lec 1, 2, 3, 4, 5
\ric$ are broadly similar. There are quantitative differences
between the present CSTF scenario for the SPP mode 1 and the
analogous CTF scenario for the sole SPP mode: the sensitivities for
$\chi_v = 15^\circ$ are greater for the latter scenario whereas the
sensitivities for $\chi_v = 60^\circ$ are greater for the former
scenario.

An alternative measure of sensitivity is provided by   the
\emph{refractive-index
 sensitivity} \cite{Homola2003}
\begin{equation}
\rho^{(\tau)}_{RI} = \frac{d \, \tilde{\theta}^{(\tau)}_{inc}
(n_\ell)}{d \, n_\ell }\,,\quad \tau\in\lec1,2,3,4,5\ric\,,
\end{equation}
also expressed in degree/RIU. While $\rho^{(\tau)}$ is the analog of
the voltage-current ratio $V/I$ in electrical circuitry,
$\rho^{(\tau)}_{RI}$ is the analog of the dynamic resistance
$dV/dI$. In Fig.~\ref{fig10},
 $\rho^{(1)}_{RI}$ and  $\rho^{(2)}_{RI}$ are plotted against $n_\ell$
 with all CSTF parameters being
 the same as those used for Fig.~\ref{fig9}.
Generally,  $\tilde{\theta}^{(\tau)}_{inc} (n_\ell)$ is
\emph{dynamically} most sensitive to changes in
 $n_\ell$ when  $n_\ell$ is large.
This general trend with respect to $n_\ell$ for the SPP mode 1
 is the {\it opposite} to that reported  for an
analogous study based on an infiltrated CTF \c{ML_PNFA}.
  The influence of $\chi_v$ on the
 dynamic sensitivity is less clear cut, as it depends upon the
 value of $n_\ell$ and which SPP mode is being considered.
The magnitudes of $\rho^{(1)}_{RI}$ and $\rho^{(2)}_{RI}$---and
$\rho^{(3)}_{RI}$,  $\rho^{(4)}_{RI}$  and $\rho^{(5)}_{RI}$ not
represented in Fig.~\ref{fig10}---are broadly similar.

\section{Closing remarks}

Numerical simulations with an empirical model have revealed that the
excitation of mutliple  SPP waves guided by the planar interface of
a metal film and a CSTF is acutely sensitive to both the refractive
index
 of a fluid infiltrating the CSTF and the morphology of the CSTF itself.
Thus, the potential for a SPP--based CSTF optical sensor has been
demonstrated. The SPP sensitivities reported here are, by and large,
similar to those reported for an analogous study based on an
infiltrated CTF \c{ML_PNFA}. However, in contrast to just one modality---angular shift
of just one SPP mode---offered by CTFs, CSTFs offer multiple modalities. These
are of two types: (i) angular shifts of more than SPP mode, and (ii) spectral shift
of
 the circular Bragg phenomenon \c{ML_IEEEPJ}. More than one of these modalities
 can be implemented simultaneously.

\vspace{10mm}

\noindent {\bf Acknowledgments:} TGM is supported by a  Royal
Academy of Engineering/Leverhulme Trust Senior Research Fellowship.
AL thanks Bernhard Michel (Simuloptics GmbH, Schwabach, Germany)
for a discussion on Reusch's 1869 paper and
 the Binder Endowment at Penn State for partial financial
support of his research activities.

\newpage
\begin{table}[ht]
\begin{center}
\begin{tabular}{|c  | c | c | c|} \hline \vspace{-4pt} &&&   \\
$\chi_v$ & $\gamma_b$ & $f$ & $n_s$   \\ &&& \vspace{-4pt} \\  \hline &&& \vspace{-4pt} \\
$15^\circ$ & 2.2793 & 0.3614 & 3.2510 \\ \hline &&& \\
$30^\circ$ & 1.8381 & 0.5039 & 3.0517 \\ \hline &&& \\
$60^\circ$ & 1.4054 & 0.6956 & 2.9105 \\ \hline
\end{tabular}
\caption{The dimensionless quantities $\gamma_b$, $f$ and $n_s$
computed  using the inverse Bruggeman homogenization formalism for a
titanium--oxide CTF with $\chi_v = 15^\circ$, $30^\circ$ and
$60^\circ$.} \label{tab1}
\end{center}
\end{table}

\newpage

\begin{figure}[!ht]
\centering
\includegraphics[width=3.5in]{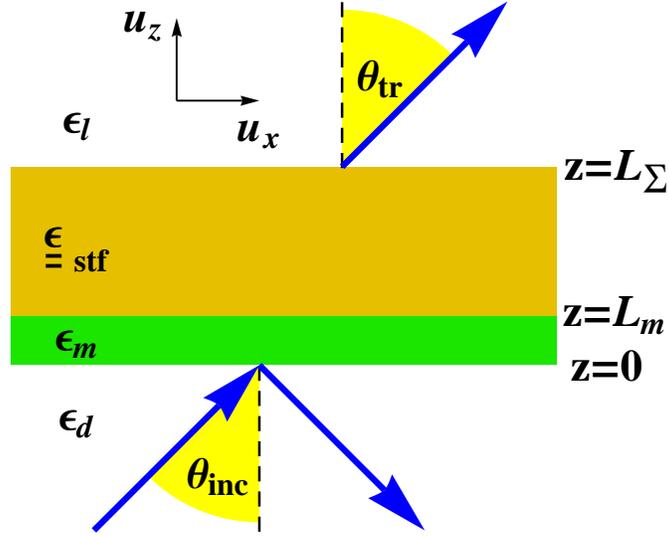}
 \caption{ \l{fig1}
A  plane wave incident at angle $\theta_{inc}$ on a metal--coated
CSTF in the modified Kretschmann configuration, giving rise to a
reflected plane wave and a transmitted plane wave.}
\end{figure}

\begin{figure}[!ht]
\centering
\includegraphics[width=2.9in]{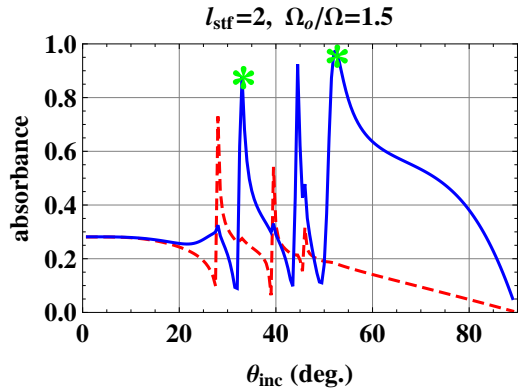}
 \caption{\l{fig2}
Absorbances for $s-$ (red, dashed  curve) and $p-$incident (blue,
solid  curve) polarization  plotted versus $\theta_{inc}$ (in
degree) for the case where $n_\ell = 1.2$, $l_{stf} = 2$, $\Omegao /
\Omega = 1.5$ and $\chi_v = 20^\circ$. The peaks corresponding the
SPP modes 1 and 2 are identified by $\star$ (in green).
  }
\end{figure}

\begin{figure}[!ht]
\centering
\includegraphics[width=2.9in]{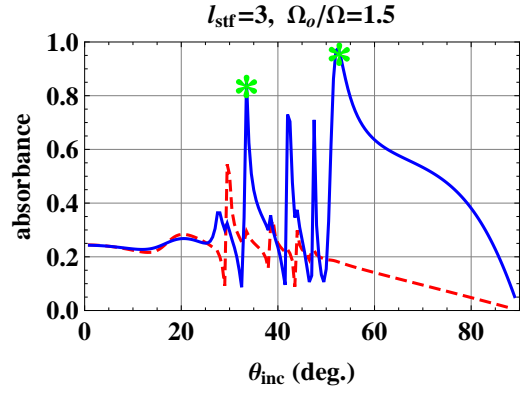}
 \caption{\l{fig3}
 As Fig.~\ref{fig2} except that $l_{stf} = 3$. }
\end{figure}

\begin{figure}[!ht]
\centering
\includegraphics[width=2.9in]{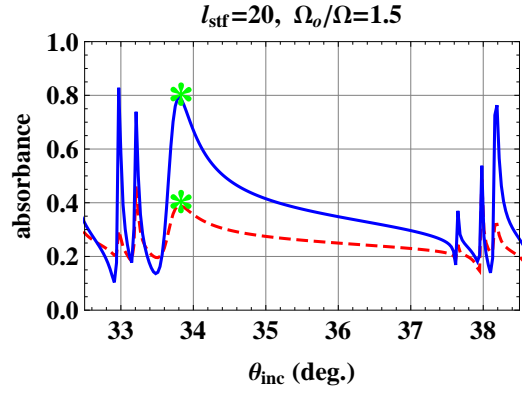}
 \caption{\l{fig4}
 As Fig.~\ref{fig2} except that $l_{stf} = 20$ and only the SPP mode 2 is represented. }
\end{figure}

\begin{figure}[!ht]
\centering
\includegraphics[width=2.9in]{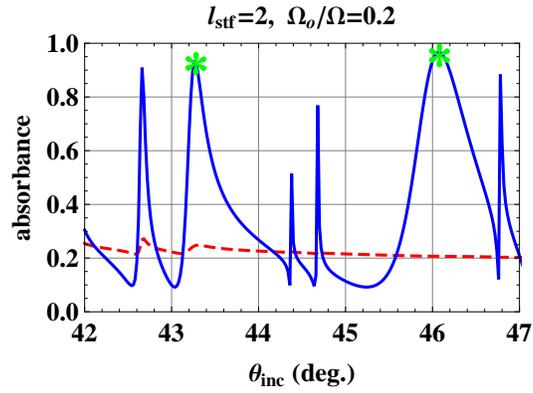}
 \caption{\l{fig5}
  As Fig.~\ref{fig2} except that $n_{\ell} = 1.1$, $\Omegao / \Omega = 0.2$ and  the SPP modes 3 and 4 are represented. }
\end{figure}

\begin{figure}[!ht]
\centering
\includegraphics[width=2.9in]{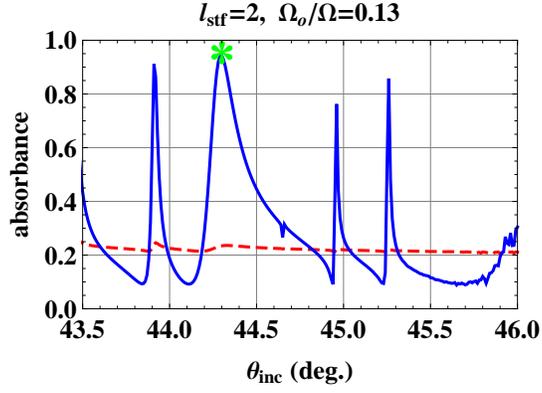}
 \caption{\l{fig6}
   As Fig.~\ref{fig2} except that $n_{\ell} = 1.1$, $\Omegao / \Omega = 0.13$ and  the SPP mode 5 is represented.  }
\end{figure}

\begin{figure}[!ht]
\centering
\includegraphics[width=2.9in]{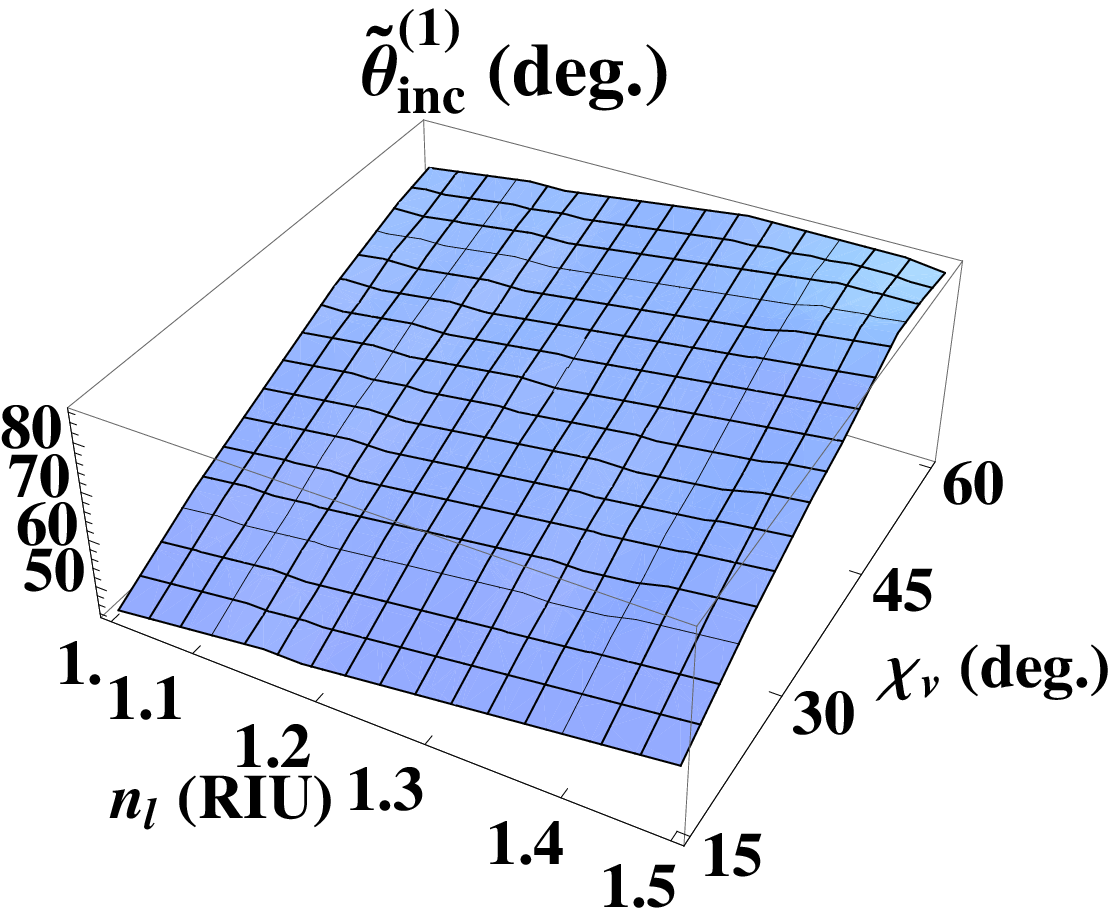}
\includegraphics[width=2.9in]{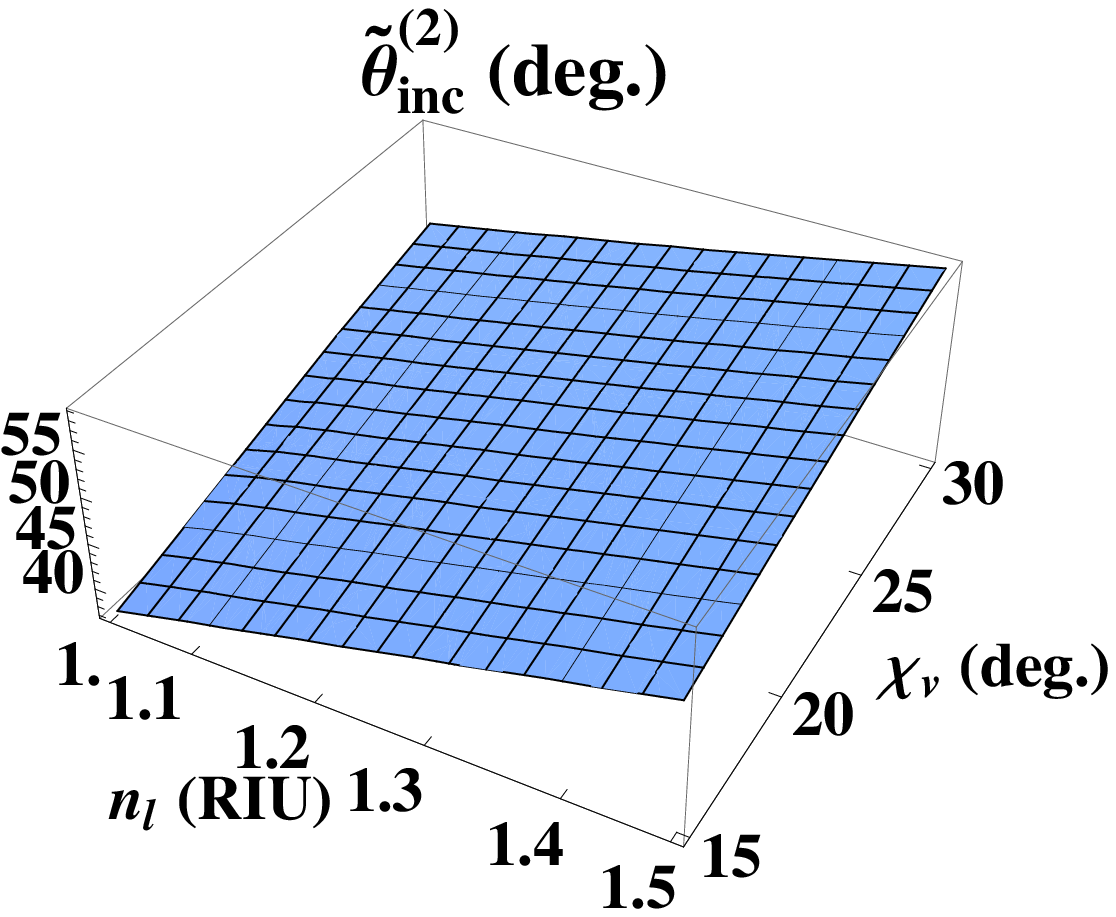}
 \caption{\l{fig7}
The angles $\tilde{\theta}^{(1)}_{inc}$ and
$\tilde{\theta}^{(2)}_{inc}$ (in degree)
  plotted versus $n_\ell$ (in RIU) and $\chi_v$ (in degree).
Here $l_{stf} = 2$ and  $\Omegao / \Omega = 0.6$.}
\end{figure}

\begin{figure}[!ht]
\centering
\includegraphics[width=2.9in]{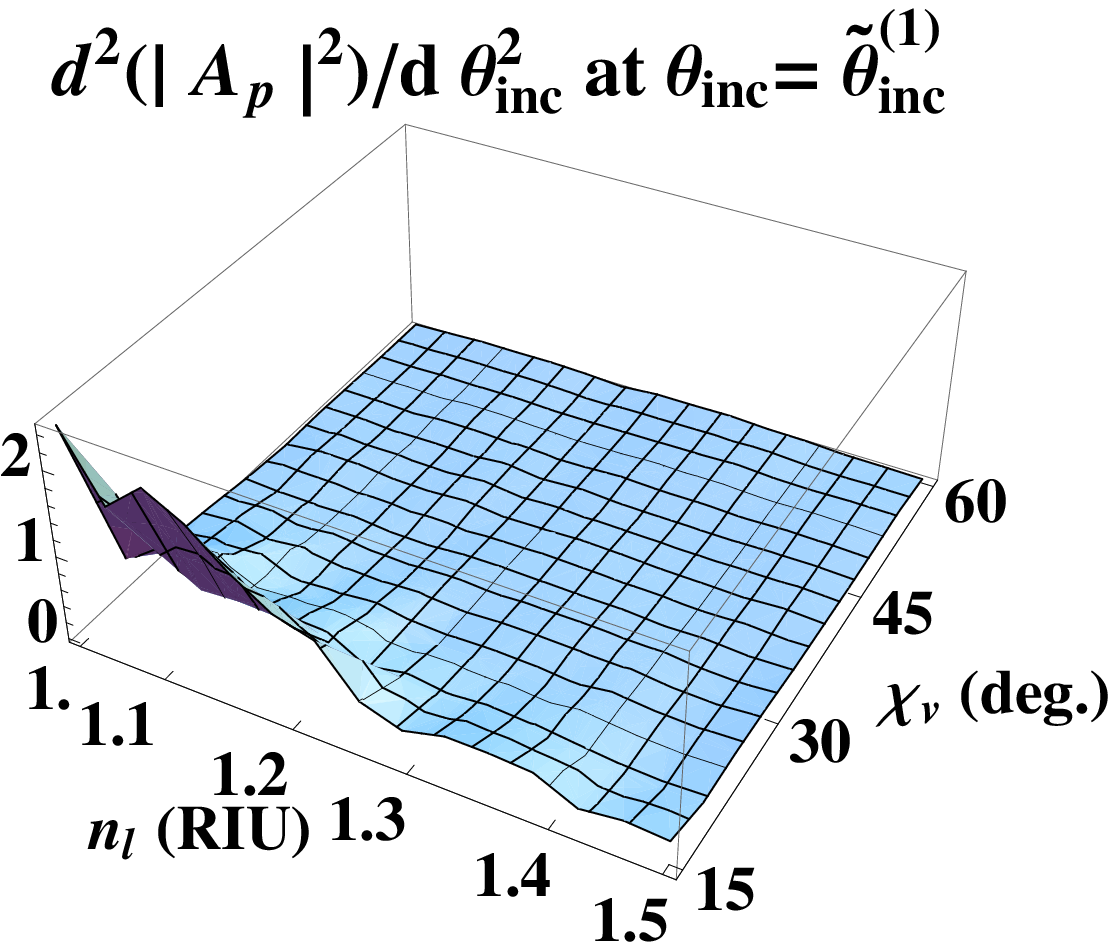}
\includegraphics[width=2.9in]{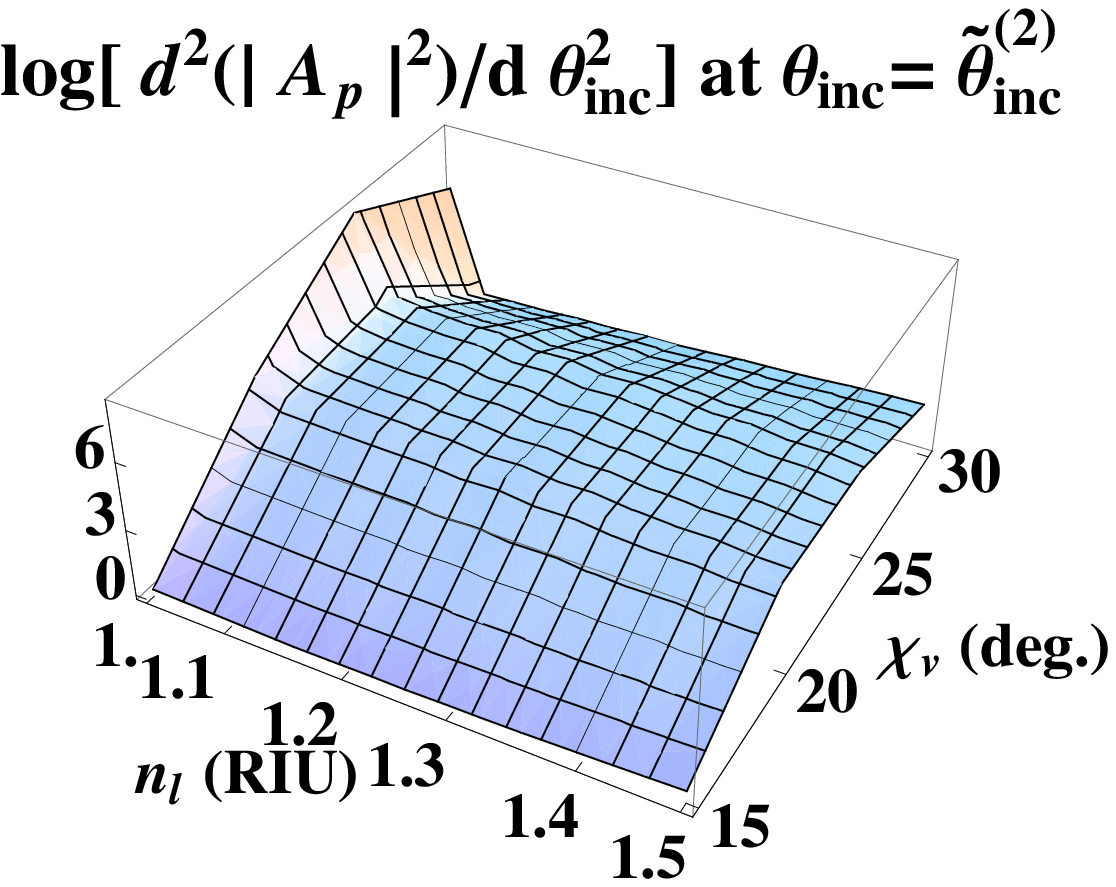}
 \caption{\l{fig8}
As Fig.~\ref{fig7} except that the quantities plotted are $d^2 \le
\left| A_p
 \right|^2 \ri / d
 \theta^2_{inc}$   (in per degree per degree)  evaluated at $\theta_{inc} = \tilde{\theta}^{(1)}_{inc}$ and
 $\log \les d^2 \le
\left| A_p
 \right|^2 \ri / d
 \theta^2_{inc} \ris$     evaluated at
  $\theta_{inc} =\tilde{\theta}^{(2)}_{inc}$.}
\end{figure}

\begin{figure}[!ht]
\centering
\includegraphics[width=2.9in]{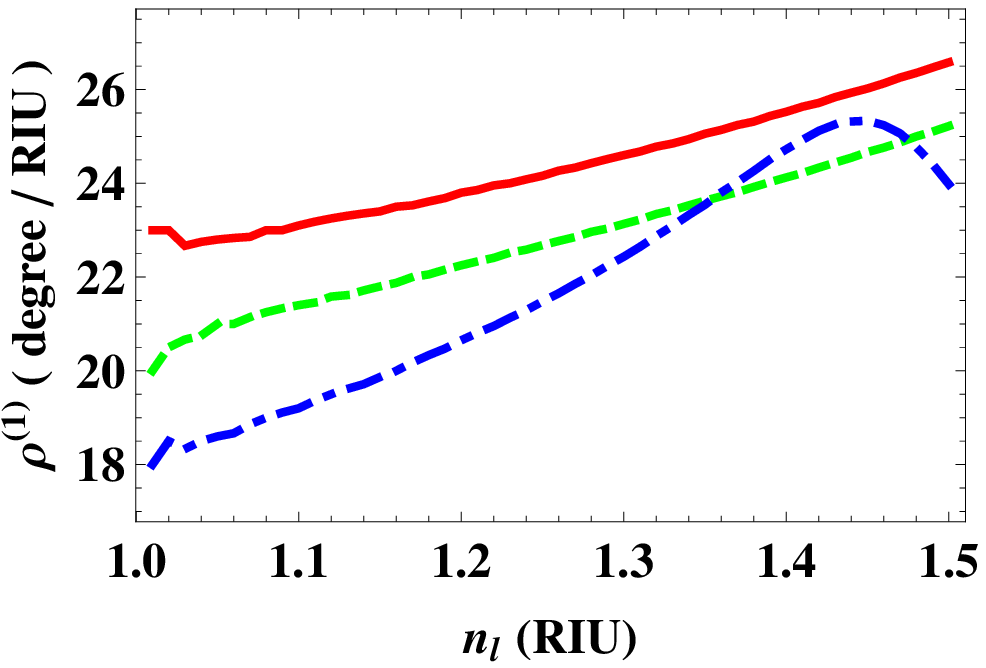}
\includegraphics[width=2.9in]{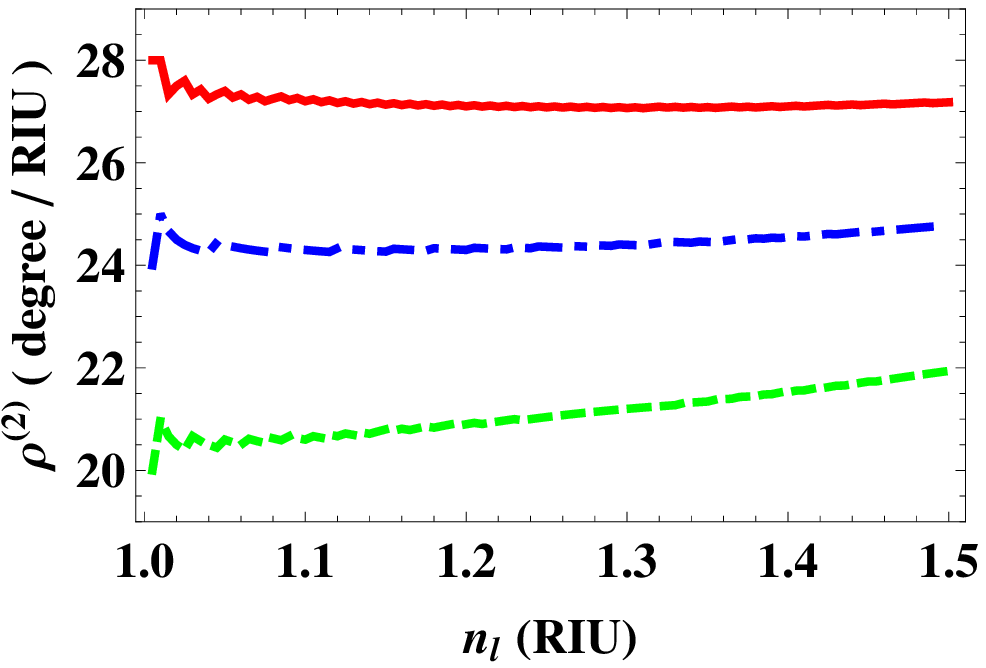}
 \caption{ \l{fig9}
The figures of merit $\rho^{(1)}$ and $\rho^{(2)}$ (in degree/RIU)
plotted against $n_\ell$ for $\chi_v = 15^\circ$ (red, thick solid
curves) and $30^\circ$ (green, dashed curves). The plots of
$\rho^{(1)}$ for $\chi_v = 60^\circ$ and $\rho^{(2)}$ for $\chi_v =
21^\circ$ are also provided  (blue, broken dashed
 curves).}
\end{figure}

\vspace{10mm}

\begin{figure}[!ht]
\centering
\includegraphics[width=2.9in]{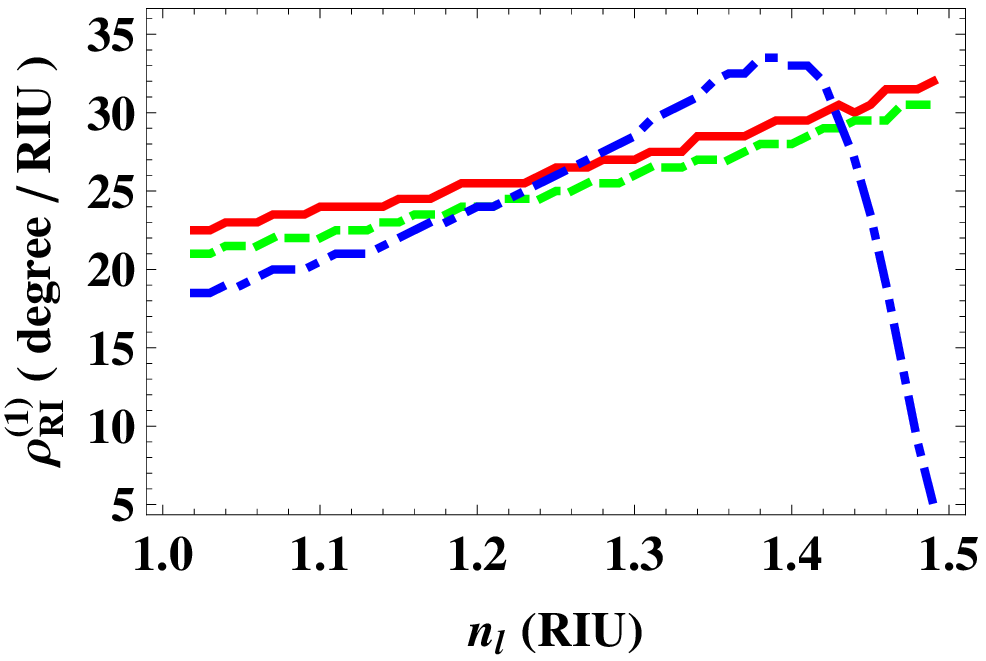}
\includegraphics[width=2.9in]{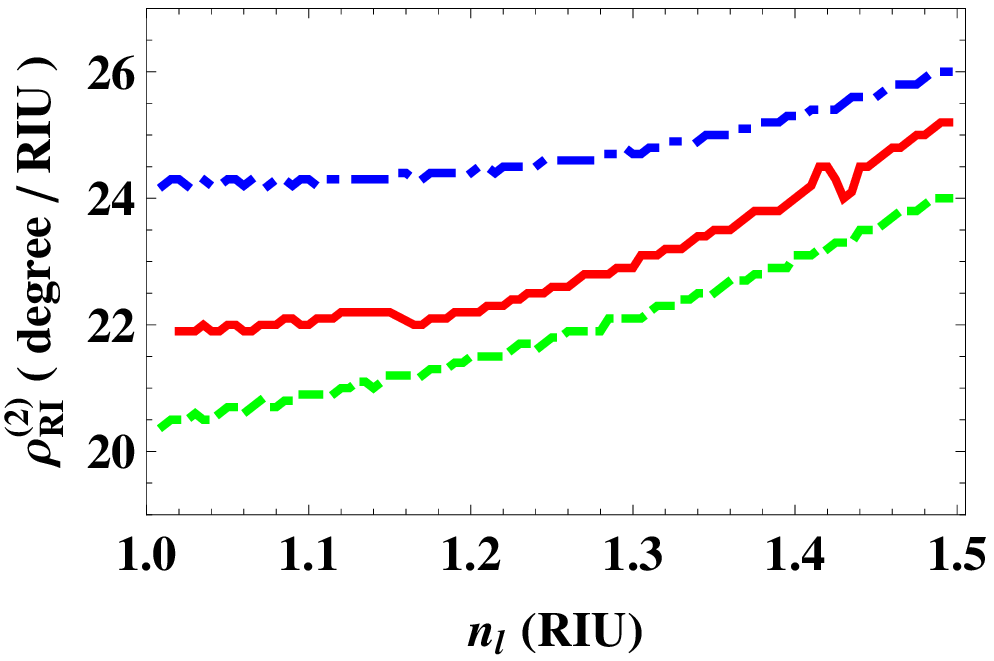}
 \caption{ \l{fig10}
As Fig.~\ref{fig9} except that the quantities plotted are the
refractive-index sensitivities $\rho^{(1)}_{RI}$ and
$\rho^{(2)}_{RI}$ (in degree/RIU).}
\end{figure}

\end{document}